# Diabatic Mechanisms of Higher-Order Harmonic Generation in Solid-State Materials under High-Intensity Electric Fields

T. Tamaya,[1*]   A. Ishikawa,[2]   T. Ogawa,[3]   and K. Tanaka[1,4,5]

[1] Department of Physics, Graduate School of Science, Kyoto University, Sakyo-ku, Kyoto 606-8502, Japan

[2] Department of Science for Advanced Materials, University of Yamanashi, 4-3-11 Takeda, Kofu, Yamanashi 400-8511, Japan

[3] Department of Physics, Osaka University, Toyonaka, Osaka 560-0043, Japan

[4] Institute for Integrated Cell-Material Sciences, Kyoto University, Sakyo-ku, Kyoto 606-8501, Japan

[5] CREST, Japan Science and Technology Agency, Kawaguchi, Saitama 332-0012, Japan

## ABSTRACT

We theoretically investigate mechanisms of higher-order harmonic generation (HHG) in solid-state materials under a high-intensity AC electric field. A new theoretical framework presented in this paper holds legitimacy of Bloch's theorem even under an influence of the high-intensity electric field and provides an exact treatment of diabatic processes of Bloch electrons. Utilizing this framework, we firstly discovered the diabatic processes, namely, AC Zener tunneling and semimetallization of semiconductors, are key factors for non-perturbative mechanisms of HHG. These mechanisms are classified by the field intensity and could be understood by extended simple man model based on an analogy between tunnel ionization in gaseous media and Zener tunneling in semiconductors. These conclusions would stimulate universal understandings of HHG mechanisms in both atomic and solid cases.

Progress in the development of intense light sources has paved the way for strong-electric-field physics and stimulated investigations on non-perturbative nonlinear optical phenomena. The most prominent matter of these investigations is higher-order harmonic generation (HHG) in gaseous media and their potential utilities for optical technology, such as attosecond pulse generation and molecular orbit tomography, have been explored [1-3]. In recent years, HHG in solid-state materials have been experimentally observed focusing on characteristics different from those in atomic cases [4-8]. The differences are based on periodic arrangement of atoms and collective properties of electrons in solid-state materials. Actually, these experiments show a band-gap dependence of a cutoff energy in HHG spectra [4,5], whose definition is a threshold between a constant intensity region (plateau region) and strong decay region, while, in atomic cases, that is determined by the relation with ionization energy of a single atom. Thus, HHG in solid-state materials is expected to have a different mechanism from that in atomic cases, and a clear understanding of such mechanism will provide a possibility for opening new research fields of high-intensity optical technology.

For the clear understanding of mechanisms of HHG in solid-state materials, we should develop an exact treatment on dynamics of the collective electrons under a high-intensity electric field. In solid-state materials where crystalline structure ensures a periodicity of atomic potential, Bloch's theorem enables us to reduce the collective behavior of electrons to single quasiparticle, which is called Bloch electron [9]. Based on this scheme, we expect an exact investigation of dynamics of Bloch electrons under the high-intensity AC electric field will unravel mechanisms of HHG. The high-intensity AC electric field causes a concurrence of excitation and transport processes of Bloch electrons, and therefore, simultaneous treatment of these processes is necessary whose considerations enable discussions on diabatic processes of Bloch electrons such as Zener tunneling [10]. However, in spite of the simplicity of this problem, previous studies [5,11-17] have not provided reasonable treatments yet because of their invalid assumptions in constructing frameworks which would be summarized as follows.

In constructing the frameworks, we should take care of legitimacy of Bloch's theorem under an influence of the high-intensity electric field. If the electric field is introduced in the form of a nonperiodic scalar potential $\phi = -e\boldsymbol{E}(t) \cdot \boldsymbol{x}$, the periodicity of atomic potential would be damaged and justification of Bloch's theorem is no longer expected to be adequate. Here, $e$ is the electron charge, $\boldsymbol{E}(t)$ the external electric field, and $\boldsymbol{x}$ the position of the electrons. In this case, Bloch functions cannot be regarded as a complete basis set and concepts of band structures and Bloch electrons become invalid [18, 19]. These invalidities would greatly affected theoretical treatments of diabatic processes, as referred in the previous work [18, 19], and their influence on perspective of HHG becomes emphasized especially when employing the high-intensity AC electric field. In spite of these failures, some previous works [5, 11, 12], where transport processes are introduced by the scalar potential, constructed their theoretical frameworks based on assumptions of the completeness of Bloch functions as well as the concepts of band structures and Bloch electrons. For correct perspective of HHG in Bloch electron picture, it is essential to hold the periodicity of atomic potential even under an influence of the high-intensity electric field.

In the simultaneous treatments of excitation and transport processes of Bloch electrons, we also should take care of justification of classical kinetic equation $\hbar\dot{\boldsymbol{k}}(t) = -e\boldsymbol{E}(t)$, where $\hbar$ and $\boldsymbol{k}$ is reduced Planck constant and Bloch wavevector. This equation is usually assumed in the treatments of transport processes of Bloch electrons and can be employed in theoretical frameworks by simple replacement of Bloch wavevector $\boldsymbol{k}$ with $\boldsymbol{k} - (e/\hbar c)\boldsymbol{A}(t)$, where the electric field is described as $\boldsymbol{E}(t) = -(1/c)\partial\boldsymbol{A}(t)/\partial t$. Here, $c$ and $\boldsymbol{A}(t)$ are the velocity of light and a vector potential. For the simultaneous treatments of excitation and transport processes, the previous works [13-16] utilized this replacement in order to introduce transport processes of Bloch electrons, while excitation processes is involved in usual dipole transition. However, as indicated by several researchers [20, 21], this replacement is true only for single-band processes and should be violated by excitation processes such as Rabi flopping [22]. In the present situation, Bloch electron states are always described as a superposition of ground and excited states, and consequently, we should conclude that the previous works could not provide an accurate perspective of HHG due to an inaccurate expression of non-perturbative interaction of Bloch electrons. Moreover, the physical interpretation of HHG mechanism based on Bloch oscillations [4, 5, 17], whose concept is true only for single-band processes, is no longer expected to be adequate. For correct perspective, the theoretical frameworks should be constructed without assuming the conventional replacement of Bloch wavevector.

In this paper, avoiding invalid assumptions employed in the previous works [5, 11-17], we clarify non-perturbative mechanisms of HHG in solid-state materials. To treat the competition between excitation and transport processes of Bloch electrons accurately, our theoretical framework started from the Hamiltonian $H = (1/2m_0)(\boldsymbol{p} - e/c\,\boldsymbol{A}(t))^2 + \Sigma_i V(x - R_i)$, where Bloch's theorem has not been applied yet. Here, $m_0$ is the electron mass, $\boldsymbol{p}$ the momentum of bare electrons, and $V(x - R_i)$ the periodic core potential of atoms located at $R_i$. In this model, we employed a homogeneous electric field in the form of a vector potential rather than a scalar one, as has been proposed by Krieger and Iafrate in the case of DC electric field [19]. An advantage of this treatment is that the Hamiltonian maintains its periodicity even including an influence of the high-intensity electric field. This ensures the validity of Bloch's theorem under the high-intensity electric field, and consequently, Bloch functions are expected to be a complete basis set and concepts of band structures and Bloch electrons become adequate. Legitimacy of Bloch's theorem including the high-intensity electric field enables us to expect that there will be temporally changed band structures linked to an AC electric field, whereas Bloch wavevector $\boldsymbol{k}$ will not change and be a kinetic constant [21]. We should take care of the Hamiltonian not always leading to the classical kinetic equation for Bloch wavevector $\hbar\dot{\boldsymbol{k}}(t) = -e\boldsymbol{E}(t)$, especially when treating excitation and transport processes together. Based on this scheme, we firstly discovered diabatic processes, namely, AC Zener tunneling and semimetallization of semiconductors, determine characteristics of HHG in solid-state materials. These considerations would provide new understandings of the fundamental mechanisms of HHG in semiconductors.

In constructing our theoretical framework, we suppose two-dimensional semiconductors which are referred in the recent experiment [5], while conclusions presented in this paper have no dependence on the dimensionality. Starting from this assumption and focusing only on conduction and valence bands, we could introduce a Hamiltonian described by $H = H_0 + H_I$ [23], where

$$H_0 = \Sigma_k \big( (E_k^e + E_g/2) \, e_k^\dagger e_k + (E_k^h + E_g/2) \, h_{-k}^\dagger h_{-k} \big), \tag{1}$$

$$H_I = \hbar\Omega_R(t)\Sigma_k \cos\theta_k \left( e_k^\dagger e_k + h_{-k}^\dagger h_{-k} - 1 \right) + i\hbar\Omega_R(t)\Sigma_k \sin\theta_k \left( e_k^\dagger h_{-k}^\dagger - h_{-k} e_k \right). \tag{2}$$

Here, $E_k^\sigma = \hbar^2 k^2/2m_\sigma (\sigma = e, h)$ are the kinetic energies of electrons and holes, $E_g$ is the band-gap energy, $e_k(h_k)$ and $e_k^\dagger(h_k^\dagger)$ are annihilation and creation operators of electrons (holes), $\theta_k$ is the argument of Bloch wavevector $\boldsymbol{k}$, and $\Omega_R(t) = \Omega_{R0} \exp(-(t-t_0)^2/\tau^2) \cos(\omega_0 t)$ is the Rabi frequency [28] where the intensity of the electric field is renormalized in $\Omega_{R0}$. Throughout this paper, we will fix the parameters of the incident electric field as $t_0 = 12\pi/\omega_0$ and $\tau = 4\pi/\omega_0$. The first term on the right hand side in Eq. (2) indicates the intraband transition in which Bloch wavevector $\boldsymbol{k}$ is a kinetic constant and it can be renormalized in the single-particle energy $\epsilon_k^\sigma$, where $\epsilon_k^\sigma(t) = E_k^\sigma + E_g/2 + \hbar\Omega_R(t) \cos\theta_k$. These modifications caused by an AC electric field mean a temporal variation of the band structure, which is consistent with the above discussion. The variations can be derived from temporal changes of diagonal matrix elements originating from the light-matter interaction that is usually ignored in nonlinear optics [22, 28]. Moreover, comparing with Landau-Zener model [29], we can interpret the Hamiltonian to include diabatic processes such as above-threshold ionization and Zener tunneling [10]. The second term of the right side in Eq. (2) indicates a dipole transition causing multiphoton absorption and Rabi flopping, which are nonlinear optical phenomena [22, 28]. The factors $\sin\theta_k$ and $\cos\theta_k$ are derived from the form factor reflecting microscopic information on the arrangement of atoms in crystals.

Considering the above-mentioned Hamiltonian, we can derive time evolution equations for populations $f_k^\sigma = \langle \sigma_k^\dagger \sigma_k \rangle$ and polarization $P_k = \langle h_{-k}^\dagger e_k \rangle$ with Bloch wavevector $\boldsymbol{k}$ as

$$i\frac{\partial}{\partial t} P_k = \left[ \epsilon_k^e(t) + \epsilon_k^h(t) \right] P_k + i\Omega_R(t) \sin\theta_k \left[ 1 - f_k^e - f_k^h \right] - i\gamma_t P_k, \tag{3}$$

$$\frac{\partial}{\partial t} f_k^\sigma = 2\text{Im}\, i\Omega_R(t) \sin\theta_k \, P_k^\dagger - \gamma_l f_k^\sigma. \tag{4}$$

Here, $\gamma_t$ and $\gamma_l$ are the transverse and longitudinal relaxation constants, and throughout this study, they will be fixed to $\gamma_t = 0.1\omega_0$ and $\gamma_l = 0.01\omega_0$, respectively. These values are estimated by the recent experimental situation [5] where the incident frequency are several tens of terahertz, while the dephasing and the energy relaxation time are regarded as a few tens of femtoseconds and a few picoseconds, respectively. The numerical solutions of these equations give the time evolutions of distributions of the carrier densities and polarization in two-dimensional $\boldsymbol{k}$ space. The distributions arise from a dipole transition under the anisotropic modification of the band structure characterized by the factor $\hbar\Omega_R(t)\cos\theta_k$, and consequently, they show anisotropic behavior in $\boldsymbol{k}$ space linked to the direction of the electric field [23]. Therefore, we expect that the intraband processes lead to carrier transport (wave packet dynamics), and the time evolution of the current can be calculated using the definition $J(t) = -c\langle\partial H_I/\partial A\rangle = \sum_k\left[\left(1 - f_k^e - f_k^h\right)\cos\theta_k - 2\text{Im}(P_k)\sin\theta_k\right]$. Accordingly, we can derive higher-order harmonic spectra from the definition $I(\omega) = |\omega J(\omega)|^2$, where $J(\omega)$ is the Fourier transform of $J(t)$, and discuss the dependence of higher-order harmonic spectra on the intensity of the incident electric field corresponding to the Rabi frequency $\Omega_{R0}$.

The numerical results show characteristics of HHG are changed depending on the Rabi frequency. In this study, we always assume a situation where $\omega_0 \ll E_g/\hbar$. At the beginning, in multiphoton absorption regime where a perturbative treatment of light-matter interaction is assured (FIG. 1(a)), we could identify the well-known characteristics in HHG spectra (FIG. 2(a)), i.e., the conventional relation of nonlinear optics $I_N \propto |P_N|^2 \propto |E_0|^{2N}$ [22] and strongly generation of only odd-order harmonics which is explained by HHG emitted in each half-cycle period of the incident electric field [1]. With increasing the Rabi frequency, unconventional non-perturbative mechanisms of HHG should be emphasized, as shown in FIG. 1(b) and FIG. 1(c), where excitation processes are dominated by Zener tunneling and semimetallization processes, respectively. To understand these regimes easily, in these figures, we assume uniform modifications of band structures described by $\epsilon_k^\sigma(t) = E_k^\sigma + E_g/2 + \hbar\Omega_R(t)$, while the anisotropic factor $\cos\theta_k$ only causes dynamics of the wave packet in $\boldsymbol{k}$ space. The threshold of the field intensity between multiphoton absorption regime and AC Zener regime can be found as the boundary when a perturbative theory becomes inadequate, which is derived as $\Omega_{R0} = 0.5\omega_0$ [30], while that between AC Zener regime and semimetal regime can be estimated as the boundary when the sum of the renormalized single-particle energy is zero: $\epsilon_{k=0}^e + \epsilon_{k=0}^h = 0$, i.e., $E_g/2\hbar = \Omega_{R0}$. In the following, we will discuss characteristics and mechanisms of HHG in each regime.

FIG. 2(b) shows the higher-order harmonic spectra in AC Zener regime, where the band-gap energies are $E_g = 10\hbar\omega_0$ (blue line) and $E_g = 5\hbar\omega_0$ (red line) in the case of $\Omega_{R0} = 2\omega_0$. We find that the spectra are divided into plateau and decay regions, and the plateau region becomes broader with increasing band-gap energy. This mechanism can be understood by an analogy between Zener tunneling and tunnel ionization processes in gaseous media [1]. The threshold between the plateau and decay regions is called cutoff energy, which corresponds to maximum energy of electron-hole pair accelerated by electric field after the pair excitation. We will consider a simple man model [1, 31] in semiconductors, as shown in FIG. 3(a), and estimate the cutoff energy $E_C$ as $E_C = 2\times(\tilde{E}_g/2 + \hbar\Omega_{R0} + 3.2U_p) = E_g + 2\times3.2U_p$, where $\tilde{E}_g = E_g - 2\hbar\Omega_{R0}$ is renormalized band-gap energy modulated by the effect of intraband transition and $U_p$ is a ponderomotive energy which represents the quiver energy of a free particle averaged over one cycle. Here, the factor 2 is derived from the two kinds of particles (electrons and holes). In addition, considering the correspondence of AC Stark shift and ponderomotive shift, we can introduce the relation $|U_p| = (1/4)\hbar\Omega_{R0}$ [30], and then the cutoff energy can be derived as $E_C \approx E_g + 1.6\hbar\Omega_{R0}$. This equation reveals that the cutoff energy shifts toward higher-energy side, in short, the plateau region becomes broader, as the bang-gap energy becomes larger. Assuming an experimental situation where an AC electric field of 30 THz is applied to GaSe and the perturbative treatments are becoming inadequate, we can expect the Rabi-frequency and band-gap energy to be $\Omega_{R0} \approx 0.5\omega_0$ and $E_g \approx 16\hbar\omega_0$. Accordingly, we can conclude that the cutoff energy is roughly determined by the band-gap energy, which is consistent with recent experimental results [5].

FIG. 2(c) shows the higher-order harmonic spectra in semimetal regime, where the band-gap energies are $E_g = 10\hbar\omega_0$ (blue line) and $E_g = 5\hbar\omega_0$ (red line) in the case of $\Omega_{R0} = 8\omega_0$. Different from multiphoton absorption and AC Zener regimes, the higher-order harmonics show a characteristic of white spectrum which masks the odd-order harmonics, and moreover, the cutoff energy no longer depends on the band-gap energy. The generation of the white higher-order harmonics means the emission law in each half cycle is broken. This half-cycle emission law can also be interpreted as temporal interference of HHG waves emitted dependently in a half cycle of the incident electric field [32, 33]. On the basis of this idea, the breaking of the emission law is due to the overlap between the conduction and valence bands by the semimetallization for a long interval, which disturbs the interference that generates only odd-order harmonic spectra. To explain the independences of the cutoff energy from the band-gap energy, we extend the simple man model, as shown in FIG. 3(b). In this model, carriers are generated when the band gap becomes zero, after which they are accelerated and finally recombine. From such processes, we can derive the cutoff energy as $E_C = 2\times(\hbar\Omega_{R0} + 3.2U_p) \approx 3.6\hbar\Omega_{R0}$, and this equation reveals that the cutoff energy no longer depends on the band-gap energy. These considerations are based on simple two-band models and their justification is assured by a condition that the band-gap energy is much smaller than energy differences between the first and the second conduction bands. Therefore, our results would be identified in experiments with several materials such as GaAs and InN.

FIGs. 4 (a) and (b) show the dependence of the cutoff frequency $\omega_C$ on the Rabi frequency, as estimated from the numerical results (purple solid circles) and the simple man model (blue and red lines) in the cases of (a) $E_g = 5\hbar\omega_0$ and (b) $E_g = 10\hbar\omega_0$. In the numerical calculation, the cutoff frequency $\omega_C$ are determined as the threshold energy between the plateau and decay regions, where the plateau region is defined from a condition that the harmonic intensities become less than 10% compared with the previous and later orders. We found a crossover between AC Zener and semimetal regimes from the dependence of the cutoff energy on the Rabi frequency.

Finally, we will clarify dephasing effect on HHG spectra whose importance is discussed in the previous works [13, 14]. The increasing dephasing effects enable us to presume disappearance of temporal interference of HHG waves, each of which is emitted dependently in a half cycle. Therefore, when employing the dephasing value of $\gamma_t \approx \omega_0$, we expect HHG spectra to be characterized by the single-cycle excitation processes, and consequently, the spectra widths become broader [34]. We identified this conjecture by performing our numerical calculations. This characteristic becomes important when considering HHG in semimetal regime. In this regime, the increasing dephasing effect changes noisy contimuumlike spectra into clean odd-ordered ones. We emphasize the cutoff laws derived in this paper are not influenced by the dephasing effect as long as considering a range of $0 \leq \gamma_t \leq \omega_0$, which covers recent experimental situations [4-8].

In this paper, based on an exact treatment of diabatic processes of Bloch electrons, we firstly discovered non-perturbative mechanisms of HHG in solid-state materials. In our theoretical framework, employing an external electric field in the form of a vector potential rather than the scalar one, justification of Bloch's theorem is ensured even under an influence of the high-intensity electric field. In this scheme, temporally changed band structures linked to an AC electric field characterize diabatic processes of Bloch electrons. Utilizing this framework, we revealed these diabatic processes, namely, AC Zener tunneling and semimetallization of semiconductors, determine properties of HHG spectra whose cutoff laws are in the form of $E_C = E_g + 1.6\hbar\Omega_{R0}$ and $E_C = 3.6\hbar\Omega_{R0}$, respectively. The cutoff laws could be understood based on analogies between Zener tunneling and tunnel ionization processes, as well as semimetallization of semiconductors and over-the-barrier ionization processes [30, 35], where band-gap energy and Bloch electrons correspond to the ionization energy and bare electrons, respectively. The analogies presented in this paper would provide possibilities for universal understandings of HHG mechanisms in both atomic and solid cases, and consequently, propose availability of the same method for coherent control such as a polarization gating [2, 36].

## ACKNOWLEDEMENT

The authors thank Y. Onishi for fruitful discussions. This work was supported by a Grant-in-Aid for Scientific Research on Innovative Areas, titled 'Optical science of dynamically correlated electrons (DYCE)' (Grant No. 20104007), by a Grant-in-Aid for Scientific Research (A) (Grant No. 26247052), by a KAKENHI (No. 26287087), and by an ImPACT Program of Council for Science, Technology and Innovation (Cabinet Office, Government of Japan).

## Figure Captions

**FIG. 1 (Color online).** Schematic diagrams of HHG mechanisms depending on the electric field intensity. (a) Multiphoton absorption regime, (b) AC Zener regime, and (c) semimetal regime. Red and blue lines show band dispersion with and without modification due to an external electric field. Green lines show excitation processes of Bloch electrons.

**FIG. 2 (Color online).** Higher harmonic spectra generated from two-dimensional semiconductors: (a) Multiphoton absorption regime, (b) AC Zener regime, and (c) semimetal regime. Red and blue lines show spectra in the case of $E_g = 5\hbar\omega_0$ and $E_g = 10\hbar\omega_0$, respectively.

**FIG. 3 (Color online).** Schematic diagram of simple man models in (a) AC Zener regime and (b) semimetal regime. In these figures, the simple man models are composed of three steps: (1) generation of carriers, (2) acceleration, and (3) recombination. Main difference of these diagrams is caused by generation processes. In AC Zener regime, carriers are excited by Zener tunneling, while in semimetallization regime, carriers are excited when the band gap becomes closed.

**FIG. 4 (Color online).** Cutoff-frequency transition from AC Zener to semimetal regimes depending on Rabi frequency in the case of $E_g = 10\hbar\omega_0$ and $E_g = 5\hbar\omega_0$. Blue and red lines show cutoff laws in AC Zener and semimetallization regimes. Purple dots indicate cutoff frequency estimated from numerical results.

**FIG. 1**

(a) multiphoton absorption regime

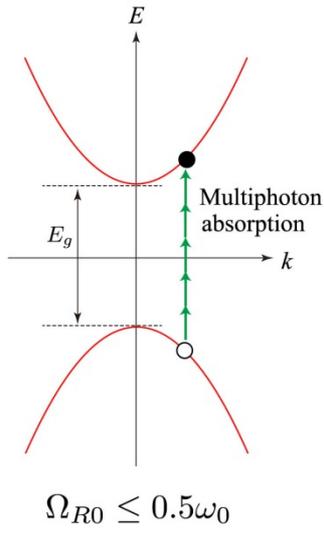

$$\Omega_{R0} \leq 0.5\omega_0$$

(b) AC Zener regime

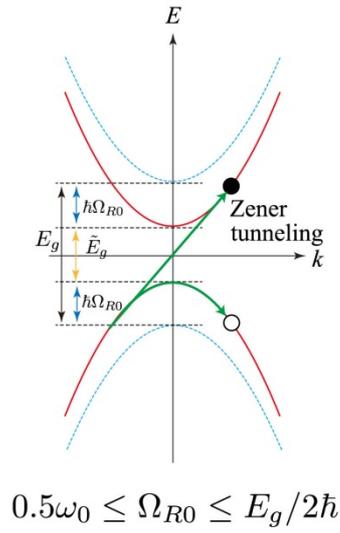

$$0.5\omega_0 \leq \Omega_{R0} \leq E_g/2\hbar$$

(c) Semimetal regime

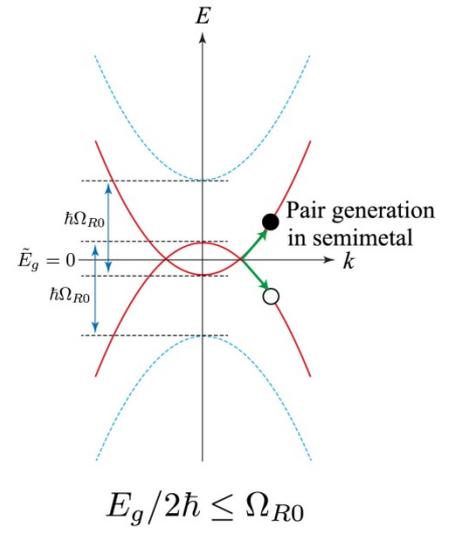

$$E_g/2\hbar \leq \Omega_{R0}$$

**T. Tamaya** *et al*

**FIG. 2**

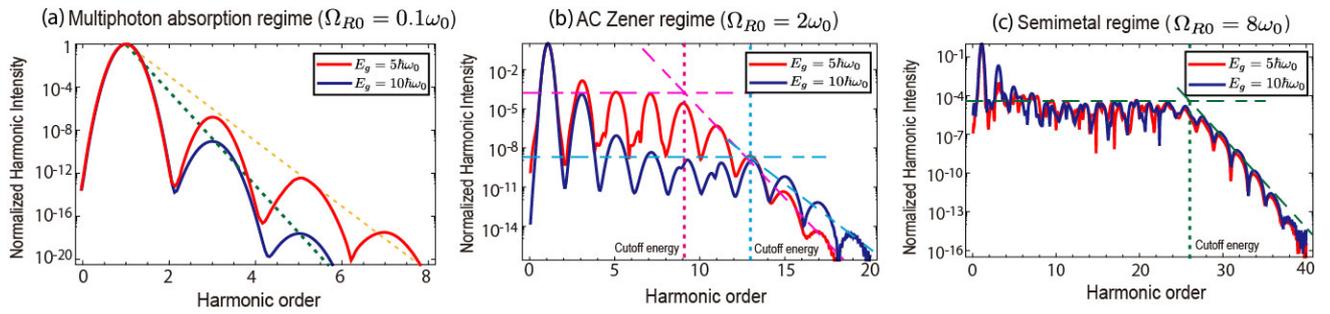

(a) Multiphoton absorption regime ($\Omega_{R0} = 0.1\omega_0$)  (b) AC Zener regime ($\Omega_{R0} = 2\omega_0$)  (c) Semimetal regime ($\Omega_{R0} = 8\omega_0$)

**T. Tamaya** *et al*

**FIG. 3**

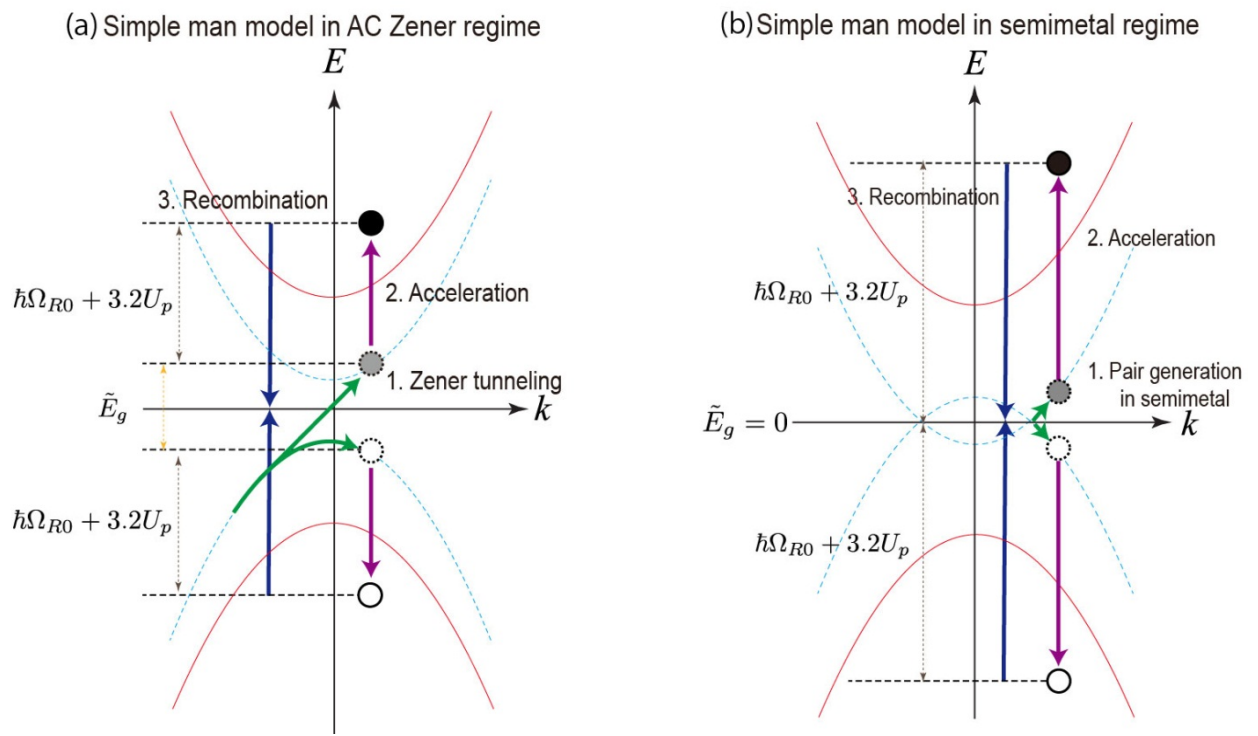

(a) Simple man model in AC Zener regime

(b) Simple man model in semimetal regime

**T. Tamaya** *et al*

**FIG. 4**

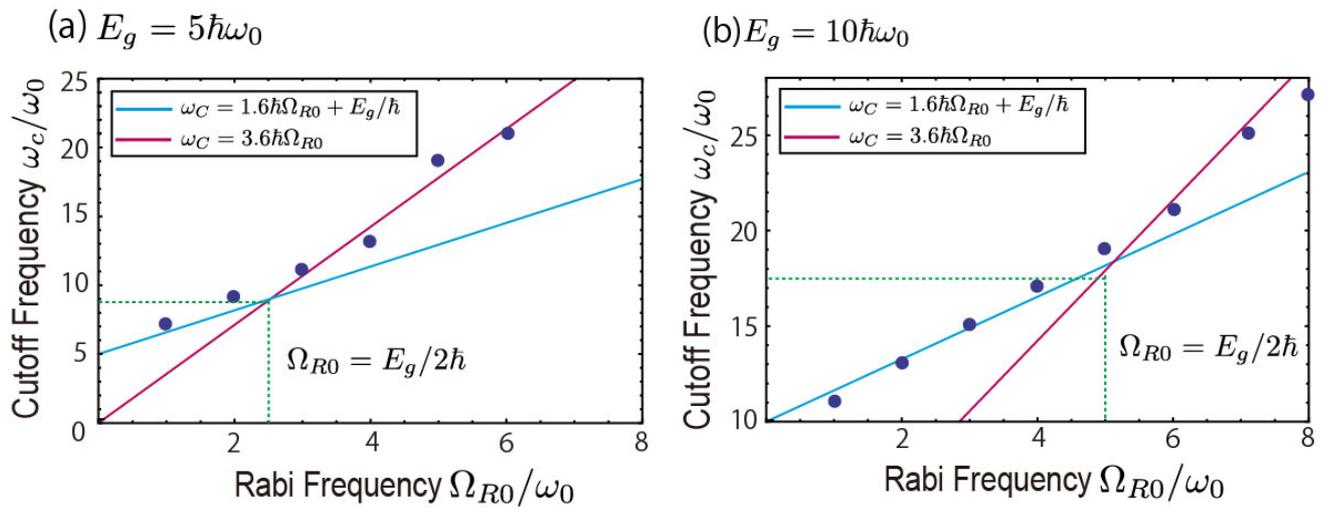

(a) $E_g = 5\hbar\omega_0$

(b) $E_g = 10\hbar\omega_0$

# Diabatic Mechanisms of Higher-Order Harmonic Generation in Solid-State Materials under High-Intensity Electric Fields: Supplementary Materials

T. Tamaya,[1*]   A. Ishikawa,[2]   T. Ogawa,[3]   and K. Tanaka[1,4,5]

[1] Department of Physics, Graduate School of Science, Kyoto University, Sakyo-ku, Kyoto 606-8502, Japan

[2] Department of Science for Advanced Materials, University of Yamanashi, 4-3-11 Takeda, Kofu, Yamanashi 400-8511, Japan

[3] Department of Physics, Osaka University, Toyonaka, Osaka 560-0043, Japan

[4] Institute for Integrated Cell-Material Sciences, Kyoto University, Sakyo-ku, Kyoto 606-8501, Japan

[5] CREST, Japan Science and Technology Agency, Kawaguchi, Saitama 332-0012, Japan

## Derivation of the Hamiltonian

To derive the Hamiltonian employed in our theory, we start from the well-known formula $H = (1/2m_0)(\boldsymbol{p} - e\boldsymbol{A}(t)/c)^2 + \Sigma_i V(x - R_i)$, where $m_0$ is the electron mass, $e$ the electron charge, $c$ the velocity of light, $\boldsymbol{A}(t)$ the vector potential of incident electric fields, $\boldsymbol{p}$ the momentum of the bare electrons, and $V(x - R_i)$ the periodic core potential of atoms located at $R_i$. Ignoring the quasi-static energy $e^2\boldsymbol{A}(t)^2/2m_0c^2$ which only shifts the total energy [1] and using the second quantization formulation, the Hamiltonian in the Coulomb gauge can be written as $H = H_0 + H_I$, where $H_0 = \int dx\, \psi^\dagger(\boldsymbol{x})\{(1/2m_0)\boldsymbol{p}^2 + \Sigma_i V(\boldsymbol{x} - R_i)\}\psi(\boldsymbol{x})$   and   $H_I = \int dx\, \psi^\dagger(\boldsymbol{x})\{-(e/m_0c)\,\boldsymbol{A}(t)\cdot\boldsymbol{p}\}\psi(\boldsymbol{x})$ . Here, $\psi(\boldsymbol{x})$ is the field operator of electrons. To consider the most basic structure of solid-state materials, we assume a two-dimensional covalent crystal in which two atoms A and B exist in a unit cell keeping the space-inversion symmetry. This assumption is equivalent to only focusing on the conduction and valence bands in a semiconductor. The incident light is assumed to be linearly polarized electric fields described as $\boldsymbol{A}(t) = \hat{\boldsymbol{x}}A(t)$. Using the tight-binding model only considering nearest-neighbor hopping of electrons and following the transformation $\hat{\psi}(\boldsymbol{x}) = (N)^{-1/2}\Sigma_{k,R_A}e^{ik\cdot R_A}\phi(\boldsymbol{x} - \boldsymbol{R_A})\hat{a}_{\boldsymbol{k}} + (N)^{-1/2}\Sigma_{k,R_B}e^{ik\cdot R_B}\phi(\boldsymbol{x} - \boldsymbol{R_B})\hat{b}_{\boldsymbol{k}}$, where the wave functions of the electrons bound to atoms A and B are described as $\phi(\boldsymbol{x} - \boldsymbol{R_A})$ and $\phi(\boldsymbol{x} - \boldsymbol{R_B})$, we can arrive at the tight-binding Hamiltonian,

$$H_0 = \Sigma_k(\gamma f(\boldsymbol{k})a_k^\dagger b_k + \gamma f^*(\boldsymbol{k})b_k^\dagger a_k), \tag{5}$$

$$H_I = -\hbar\Sigma_k(\Omega_R(\boldsymbol{k}, t)a_k^\dagger b_k + h.c.). \tag{6}$$

Here, $\gamma$ is the transfer integral, $f(\boldsymbol{k})$ the form factor $f(\boldsymbol{k}) = \Sigma_i e^{ik\cdot\delta_i} = |f(\boldsymbol{k})|e^{i\theta_{f(k)}}$, $\boldsymbol{\delta}_i$ the lattice vector, $a_k$ $(b_k)$ the annihilation operators of electrons with the wavenumber $\boldsymbol{k}$ on the sub lattice

A (B), and $\Omega_R(\boldsymbol{k}, t)$ the Rabi frequency defined by $\Omega_R(\boldsymbol{k}, t) = (e\hbar\pi(\boldsymbol{k}, t)/m_0 c)$ where

$$\pi(\boldsymbol{k}, t) = \Sigma_i e^{i\boldsymbol{k} \cdot \boldsymbol{\delta}_i} \int d^2x\, \phi^*(\boldsymbol{x}) \boldsymbol{A}(t) \cdot \boldsymbol{p} \phi(\boldsymbol{x} - \boldsymbol{\delta}_i). \qquad (7)$$

In the following formulation, we will ignore the $\boldsymbol{k}$ dependence of the Rabi frequency, which is usually permitted in semiconductor physics [2]. In this Hamiltonian, the diagonal matrix elements in Eqs. (5) and (6) become zero because of the space-inversion symmetry in a semiconductor and no carrier doping. The transformation to the band-structure picture can be performed by diagonalization of the single-particle part $H_0$ through the use of the electron-hole picture defined as $e_k = 1/\sqrt{2}\big[a_k + e^{i\theta_{f(k)}} b_k\big]$ and $h^\dagger_{-k} = 1/\sqrt{2}\big[-a_k + e^{i\theta_{f(k)}} b_k\big]$. The Hamiltonian in the electron-hole picture are derived as $H = H_0 + H_I$ , where $H_0 = \Sigma_k\big(\gamma|f(\boldsymbol{k})|\, e^\dagger_k e_k + \gamma|f(\boldsymbol{k})|\, h^\dagger_{-k} h_{-k}\big)$ and $H_I = \hbar\Omega_R(t)\Sigma_k \cos\theta_{f(k)} \big(e^\dagger_k e_k + h^\dagger_{-k} h_{-k} - 1\big) + i\hbar\Omega_R(t)\Sigma_k \sin\theta_{f(k)} \big(e^\dagger_k h^\dagger_{-k} - h_{-k} e_k\big)$. Here, $e_k(h_k)$ is the annihilation operator of electrons (holes). In this expression, we have ignored the constant energy term that gives the total-energy shift. To obtain a more specific description, we should include microscopic information on the arrangement of the atoms in the crystal. To consider the most simplified case, we assume $\gamma|f(\boldsymbol{k})| \approx (\hbar^2\boldsymbol{k}^2/2m_\sigma + E_g/2)$ and $\theta_{f(k)} = \theta_{\boldsymbol{k}}$. Thus, we can derive a Hamiltonian in the form of

$$H_0 = \Sigma_k\big((E^e_k + E_g/2)\, e^\dagger_k e_k + (E^h_k + E_g/2)\, h^\dagger_{-k} h_{-k}\big), \qquad (8)$$

$$H_I = \hbar\Omega_R(t)\Sigma_k \cos\theta_k \big(e^\dagger_k e_k + h^\dagger_{-k} h_{-k} - 1\big) + i\hbar\Omega_R(t)\Sigma_k \sin\theta_k \big(e^\dagger_k h^\dagger_{-k} - h_{-k} e_k\big). \qquad (9)$$

The similar derivation of the Hamiltonian in the electron-hole picture was performed in graphene systems, where the hexagonal lattice structure is the essential property [3, 4].

### Some characteristics of the Hamiltonian

We should take care of the differences between the derived Hamiltonian and the conventional one employed in semiconductor physics. The conventional Hamiltonian in semiconductor physics usually takes the form [5],

$$H_0 = \Sigma_{\boldsymbol{k}}\big((E^e_{\boldsymbol{k}} + E_g/2)\, e^\dagger_{\boldsymbol{k}} e_{\boldsymbol{k}} + (E^h_{\boldsymbol{k}} + E_g/2)\, h^\dagger_{-\boldsymbol{k}} h_{-\boldsymbol{k}}\big), \qquad (10)$$

$$H_I = i\hbar\Omega_R(t)\Sigma_k(e_k^\dagger h_{-k}^\dagger - h_{-k}e_k).\qquad(11)$$

Here, we can see that the main difference of Eqs. (4) and (5) from Eqs. (6) and (7) is the first term in Eq. (2) that is scaled by the Rabi frequency $\Omega_R(t)$. This term indicates the intraband transition of Bloch electrons where the Bloch wavevector $\boldsymbol{k}$ is a kinetic constant. In semiconductor physics, this process is usually omitted because of the selection rule of the dipole transition, where the band structure is assumed to be fixed even under a strong AC electric field (the rigid-band assumption) [5]. In contrast, our theory naturally includes this term which is renormalized in the single-particle energy $\epsilon_k^\sigma$ as $\epsilon_k^\sigma(t) = E_k^\sigma + E_g/2 + \hbar\Omega_R(t)\cos\theta_k$, where the band structure varies with the strength of the AC electric field. Comparing these expressions and focusing on the characteristics of the first term in Eq. (2) that is scaled by the Rabi frequency $\Omega_R(t)$, we can conclude that the rigid-band assumption is only true for a weak electric field, in short, $\hbar\Omega_R(t) \ll E_g$. Therefore, in treating the carrier dynamics under high-intensity electric fields, the modification of the band structure becomes important and leads to a violation of the selection rule, as has been pointed out in the discussions of gaseous media [6]. By comparing with the Landau-Zener model [7], we can also interpret the first term in Eq. (2) describes the diabatic processes such as the above-threshold ionization and Zener tunneling [1, 8]. Other differences between the derived and conventional Hamiltonians are the factors $\sin\theta_k$ and $\cos\theta_k$ which are derived from the form factor defined by $f(\boldsymbol{k}) = \Sigma_i e^{i\boldsymbol{k}\cdot\boldsymbol{\delta}_i}$. These factors reflect microscopic information on the arrangement of atoms in the crystal, which are never included in the conventional Hamiltonian employed in semiconductor physics. In particular, the factor $\cos\theta_k$, which characterizes the single-particle energy, $\epsilon_k^\sigma(t) = E_k^\sigma + E_g/2 + \hbar\Omega_R(t)\cos\theta_k$, indicates an anisotropic deformation of the band structure, and the dipole transition under such anisotropic modifications leads to the aspect of transport phenomena.

## Numerical results of kinetic equations

The above Hamiltonian enables us to derive time evolution equations for the populations $f_k^\sigma = \langle\sigma_k^\dagger\sigma_k\rangle$ and polarization $P_k = \langle h_{-k}^\dagger e_k\rangle$ with the Bloch wavevector $\boldsymbol{k}$ in the form of

$$i\hbar\frac{\partial}{\partial t}P_k = \left[\epsilon_k^e(t) + \epsilon_k^h(t)\right]P_k + i\Omega_R(t)\sin\theta_k\left[1 - f_k^e - f_k^h\right] - i\gamma_t P_k,\qquad(12)$$

$$\hbar\frac{\partial}{\partial t}f_k^\sigma = 2\mathrm{Im}\,i\Omega_R(t)\sin\theta_k\,P_k^\dagger - \gamma_l f_k^\sigma.\qquad(13)$$

Here, $\Omega_R(t) = \Omega_{R0}\exp(-(t-t_0)^2/\tau^2)\cos(\omega_0 t)$ is the Rabi frequency in which the intensity of the electric field is renormalized in $\Omega_{R0}$, $\epsilon_k^\sigma(t) = E_k^\sigma + E_g/2 + \hbar\Omega_R(t)\cos\theta_k$ is the single-particle energy modulated by the AC electric field, and $\gamma_t$ and $\gamma_l$ are the transverse and longitudinal relaxation constants. Throughout this study, the parameters of the incident electric field will be fixed to $t_0 = 12\pi/\omega_0$ and $\tau = 4\pi/\omega_0$ (see the top panel of FIG. 1), and the relaxation constants to $\gamma_t = 0.1\hbar\omega_0$ and $\gamma_l = 0.01\hbar\omega_0$, respectively. As discussed in the main paper, the temporal variation of the single-particle energy indicates the anisotropic modification of the band structure, and the dipole transition under such modifications leads to the anisotropic distributions of the carrier densities and polarization in two-dimensional $\boldsymbol{k}$ space. FIGs. 1(a1) and (b1) show the densities $f_k^e = f_k^h$ and (a2) and (b2) show the distributions of the polarizations $2\mathrm{Im}[P_k]$, where $E_g = 5\hbar\omega_0$ and $\Omega_{R0} = 2\omega_0$ in the case of $t = 12\pi/\omega_0$ and $t = 13\pi/\omega_0$. These figures mean that the anisotropic distributions in $\boldsymbol{k}$ space are linked to the direction of the electric field and that they arise from a dipole transition under the anisotropic modification of the band structure characterized by the factor $\hbar\Omega_R(t)\cos\theta_k$. Therefore, we expect that the intraband processes lead to carrier transport (wave packet dynamics), and the time evolution of the current can be calculated using the definition $J(t) = -c\langle\partial H_I/\partial A\rangle = \sum_k\big[(1 - f_k^e - f_k^h)\cos\theta_k - 2\mathrm{Im}(P_k)\sin\theta_k\big]$.

* ttamaya@scphys.kyoto-u.ac.jp

**FIG. 1 (Color online).** Top panel: Waveform of the incident electric field. Bottom panel: Density and polarization distributions in the case of $E_g = 5\hbar\omega_0$ and $\Omega_{R0} = 2\omega_0$ for $t = 12\pi/\omega_0$ ((a1) and (a2)) and $t = 13\pi/\omega_0$ ((b1) and (b2)).

FIG. 1

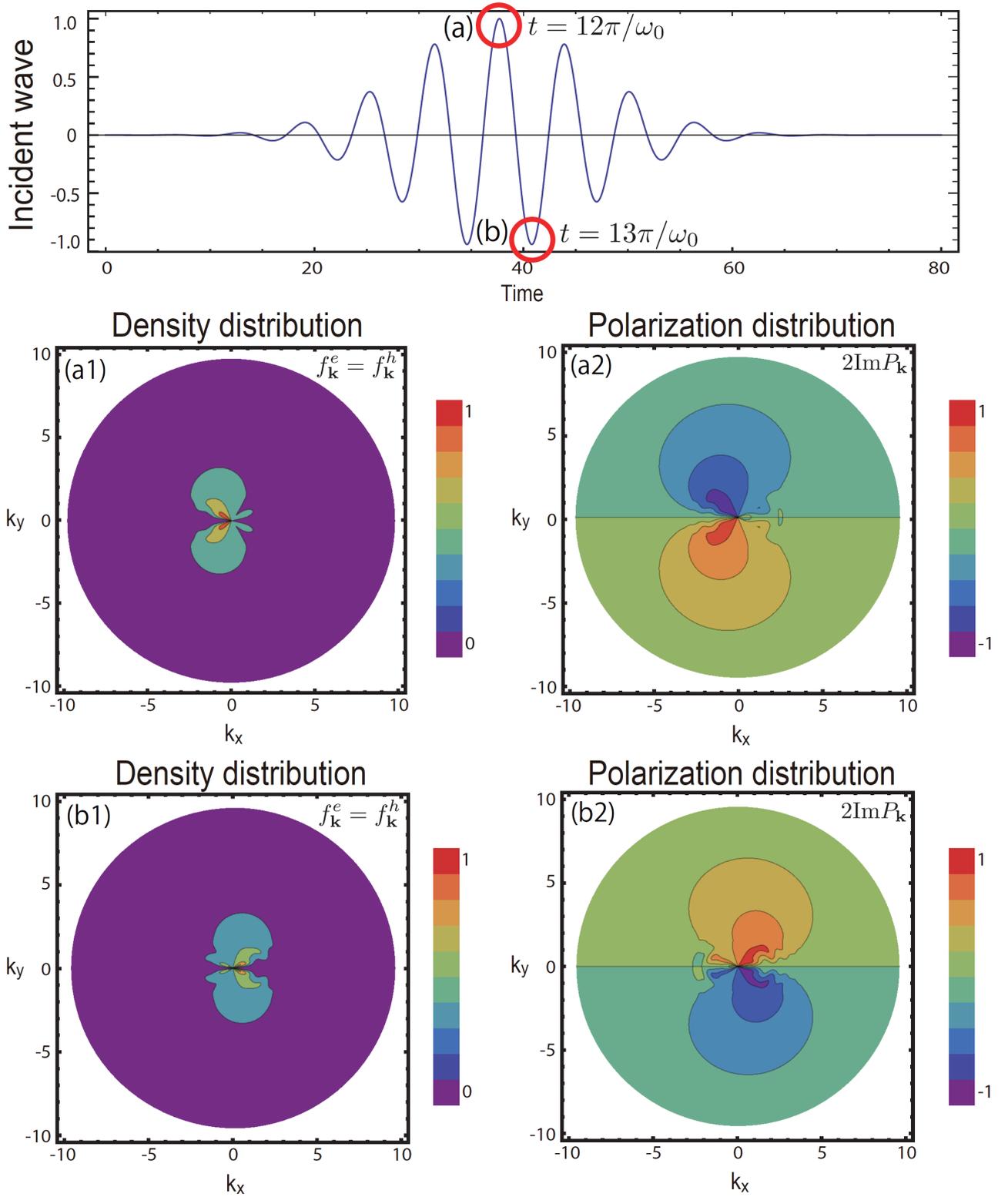